%% file: conference_101719.tex
\documentclass[conference,compsoc]{IEEEtran}
\usepackage[nocompress]{cite}
\usepackage{amsmath,amssymb,amsfonts}
\usepackage{algorithmic}
\usepackage[pdftex]{graphicx}
\usepackage{textcomp}
\usepackage{xcolor}
\usepackage[caption=false,font=footnotesize,labelfont=sf,textfont=sf]{subfig}
\usepackage{url}

\usepackage{makecell}
\usepackage{booktabs}

\def\BibTeX{{\rm B\kern-.05em{\sc i\kern-.025em b}\kern-.08em
    T\kern-.1667em\lower.7ex\hbox{E}\kern-.125emX}}

\usepackage[normalem]{ulem}

\newcommand{\bu}[1]{\textbf{\underline{#1}}}
\begin{document}

\title{SPENDER: A Platform for Secure and Privacy-Preserving \\ Decentralized P2P E-Commerce}
% \author{Only for \textit{Pre-Print}}
% \author{Anonymous Author(s)}
\author{
    \IEEEauthorblockN{Shuhao Zheng}
    \IEEEauthorblockA{McGill University}
    \and
    \IEEEauthorblockN{Junliang Luo}
    \IEEEauthorblockA{McGill University}
    \and
    \IEEEauthorblockN{Erqun Dong}
    \IEEEauthorblockA{McGill University}
   \and
    \IEEEauthorblockN{Can Chen}
    \IEEEauthorblockA{McGill University}
   \and
    \IEEEauthorblockN{Xue Liu}
    \IEEEauthorblockA{McGill University}
}
% \and
% \and
% \and
% \and
% \and

\maketitle

\input{sections/abstract}
\input{figures/front_end}
\input{sections/sec1-intro}
\input{sections/sec2-preliminary}
\input{figures/pipeline}
\input{sections/sec3-related-work}
\input{sections/sec4-sys-pipe}
\input{tables/backend_message}
\input{sections/sec5-implementation}

\input{sections/sec6-adv-features}
\input{sections/sec7-future-potentials}
\input{sections/sec8-conclusion}
% \newpage
% \bibitem{b10} R. T. Wigand, “Electronic commerce: Definition, theory, and context,”
% The information society, vol. 13, no. 1, pp. 1–16, 1997.
% \bibitem{b11}H. Treiblmaier and C. Sillaber, “The impact of blockchain on e-
% commerce: A framework for salient research topics,” Electronic Com-
% merce Research and Applications, vol. 48, p. 101054, 2021.

% \end{thebibliography}

\bibliographystyle{IEEEtran}
\bibliography{IEEEabrv,ref.bib}

\end{document}

%% file: sections/abstract.tex
\begin{abstract}
The blockchain technology empowers secure, trustless, and privacy-preserving trading with cryptocurrencies.
However, existing blockchain-based trading platforms only support trading cryptocurrencies with digital assets (e.g., NFTs).
Although several payment service providers have started to accept cryptocurrency as a payment method for tangible goods (e.g., Visa, PayPal), customers still need to trust and hand over their private information to centralized E-commerce platforms (e.g., Amazon, eBay).
To enable trustless and privacy-preserving trading between cryptocurrencies and real goods, we propose \bu{SPENDER}, a smart-contract-based platform for \bu{S}ecure and \bu{P}rivacy-Pres\bu{E}rvi\bu{N}g \bu{D}ecentralized P2P \bu{E}-comme\bu{R}ce.
The design of our platform enables various advantageous features and brings unlimited future potential.
Moreover, our platform provides a complete paradigm for designing real-world Web3 infrastructures on the blockchain, which broadens the application scope and exploits the intrinsic values of cryptocurrencies.
The platform has been built and tested on the Terra ecosystem, and we plan to open-source the code later.
\end{abstract}

% \begin{IEEEkeywords}
% Web3, Blockchain, E-commerce, Secure, Privacy
% \end{IEEEkeywords}

\IEEEpeerreviewmaketitle

%% file: figures/front_end.tex
% \begin{figure*}
%     \centering
%      \begin{subfigure}[b]{0.30\textwidth}
%      \begin{minipage}[b]{0.99\textwidth}
%      \includegraphics[width=\textwidth]{images/seller_end.png}
%      \caption{Seller's end}
%      \label{fig:buy_end}
%      \end{minipage}
%      \begin{minipage}[b]{0.99\textwidth}
%      \centering
%      \includegraphics[width=\textwidth]{images/shipper_end.png}
%      \caption{Shipper's end}
%      \label{fig:buy_end}
%      \end{minipage}
%      \end{subfigure}
%      \hfill
%      \begin{subfigure}[b]{0.59\textwidth}
%      \centering
%      \hspace{1cm}
%      \vspace{0.6cm}
%      \includegraphics[width=\textwidth]{images/buyer_end.png}
%      \caption{Buyer's end}
%      \label{fig:ship_end}
%      \end{subfigure}
%     \caption{D-Commerce Front-end}
%     \label{fig:front_end}
% \end{figure*}

\begin{figure*}
    \centering
    \subfloat[Buyer's end]{\includegraphics[width=0.32\textwidth]{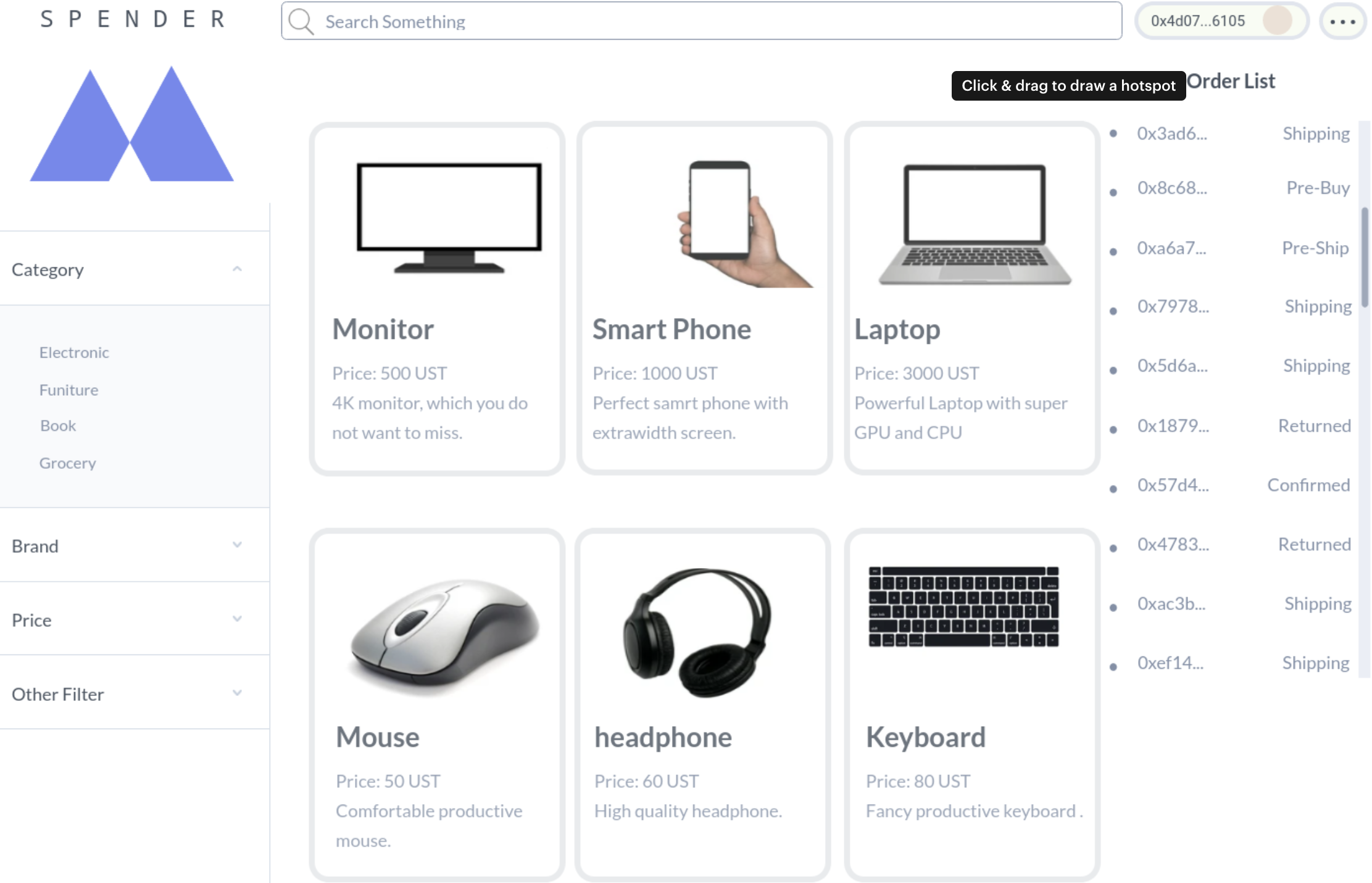}%
    \label{fig:buy_end}}
    \hfill
    \subfloat[Seller's end]{\includegraphics[width=0.32\textwidth]{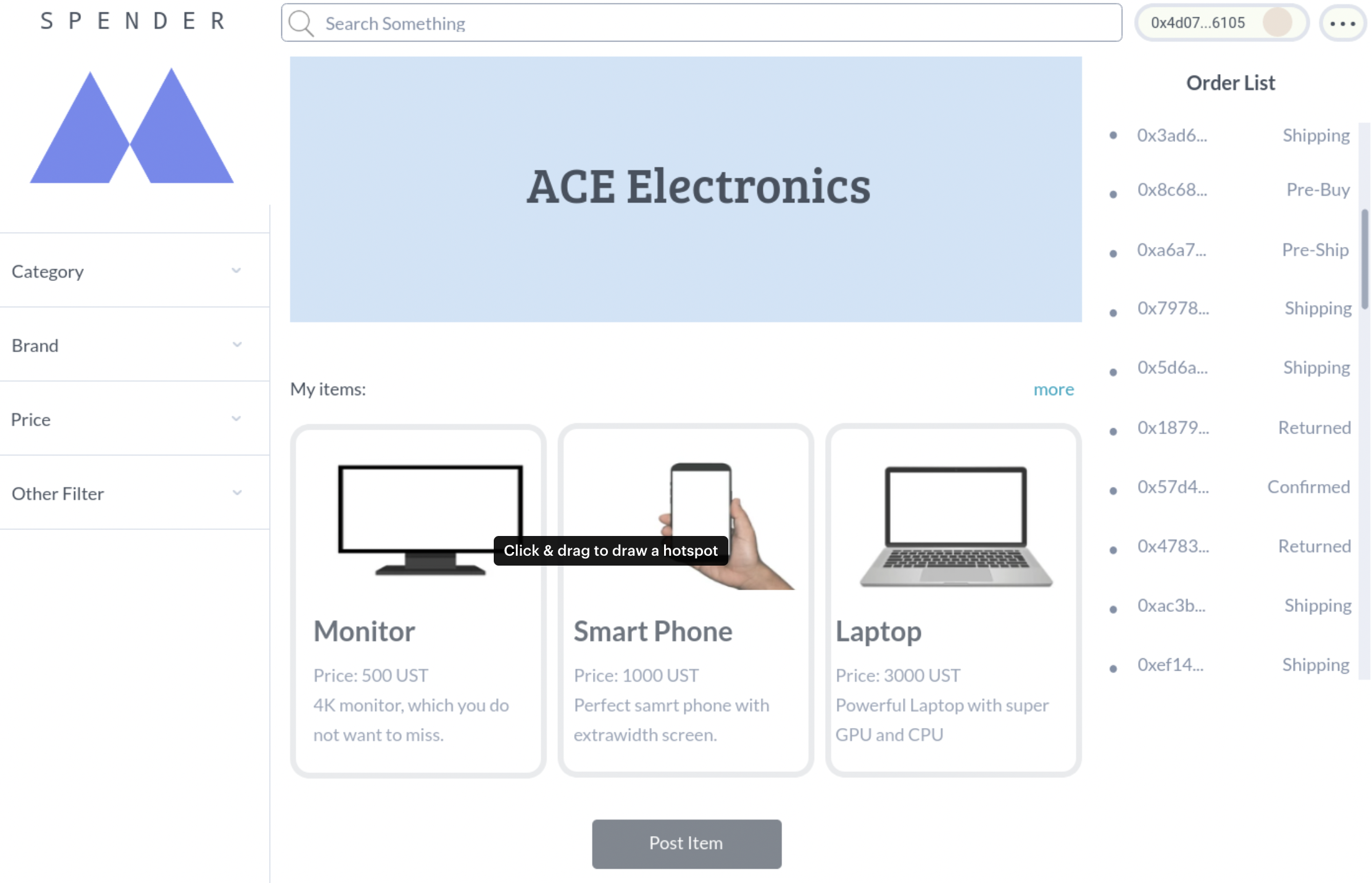}%
    \label{fig:sell_end}}
    \hfill
    \subfloat[Shipper's end]{\includegraphics[width=0.32\textwidth]{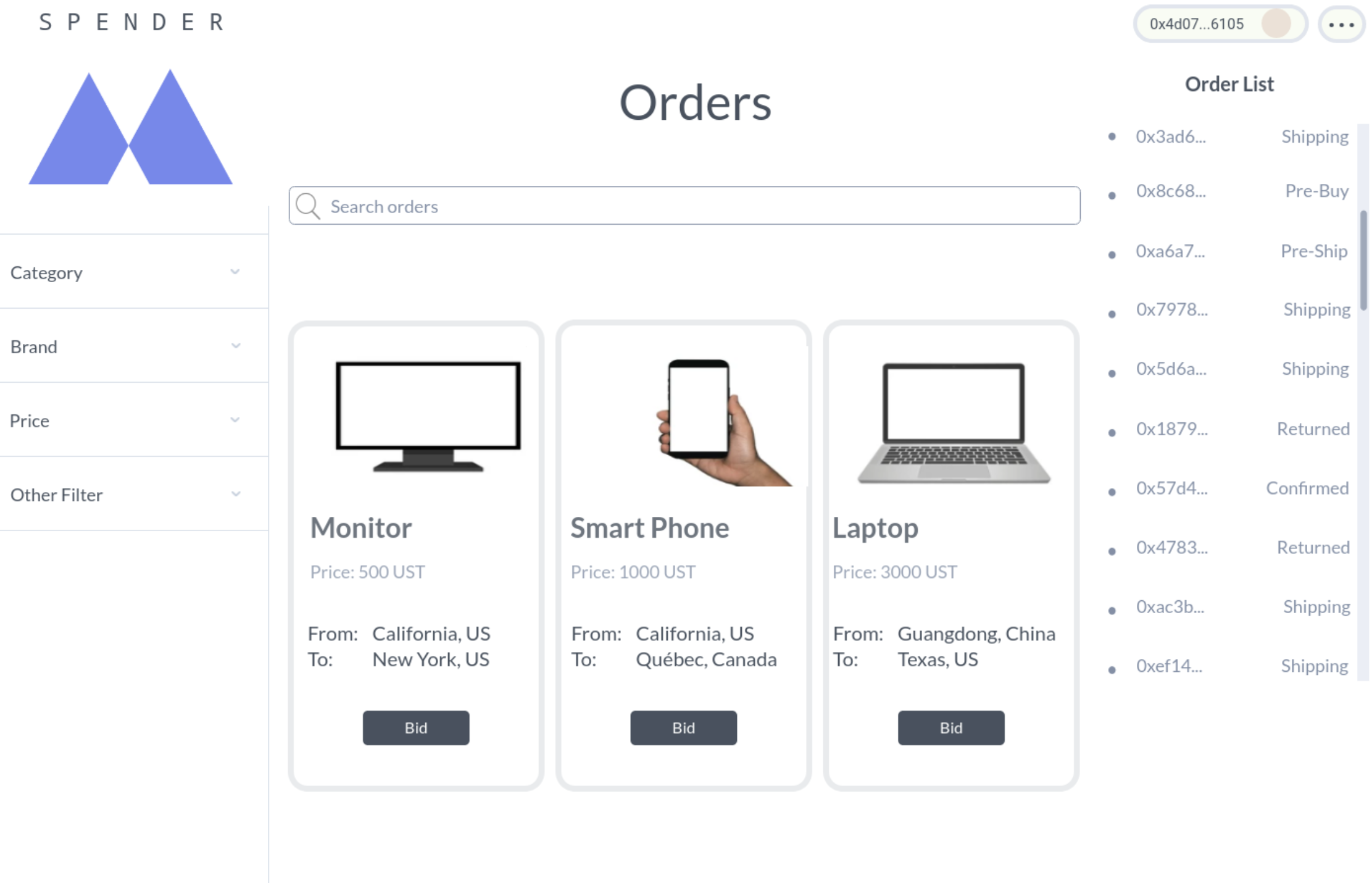}%
    \label{fig:ship_end}}
    %  \begin{subfigure}[b]{0.32\textwidth}
    %  \centering
    %  \includegraphics[width=\textwidth]{images/buyer_end.png}
    %  \caption{Buyer's end}
    %  \label{fig:buy_end}
    %  \end{subfigure}
    %  \hfill
    %  \begin{subfigure}[b]{0.32\textwidth}
    %  \centering
    %  \includegraphics[width=\textwidth]{images/seller_end.png}
    %  \caption{Seller's end}
    %  \label{fig:sell_end}
    %  \end{subfigure}
    %  \hfill
    %  \begin{subfigure}[b]{0.32\textwidth}
    %  \centering
    %  \includegraphics[width=\textwidth]{images/shipper_end.png}
    %  \caption{Shipper's end}
    %  \label{fig:ship_end}
    %  \end{subfigure}
    \caption{Graphical User Interface (GUI) Design of SPENDER Platform.}
    \label{fig:front_end}
\end{figure*}

%% file: sections/sec1-intro.tex
\section{Introduction}
The blockchain is naturally a secure environment for anonymous and trustless transactions without involving any centralized authorities \cite{nakamoto2008bitcoin}.
All these transactions are made through cryptocurrencies which serve as utility tokens.
In recent years, different blockchains have supported new features and functionalities, especially smart contracts \cite{savelyev2017contract}.
The smart contracts bolster infinite possibilities for decentralized applications (DApps) built on-chain.
Among the emerging DApps, platforms that enable trading digital assets with cryptocurrencies have been increasingly sparking interest in Web3 communities \cite{wang2021non}.
For example, OpenSea\footnote{https://opensea.io/}, the world's largest marketplace for trading Non-Fungible Tokens (NFTs), already has an NFT trading volume worth more than $\$23$ billion\footnote{https://dappradar.com/multichain/marketplaces/opensea}.
%according to DappRader
%
Other types of digital assets including digital contents \cite{li2020lbry} (e.g., videos and music), data \cite{dixit2021fast}, etc., are also being traded on blockchains.

Although digital asset trading with cryptocurrencies has been thriving, trading tangible goods with cryptocurrencies, i.e., decentralized E-commerce, has yet to be developed.
One big challenge in decentralized E-commerce is that no practical paradigm could facilitate both on-chain communications and off-chain trading interactions.
For one thing, allowing off-chain interactions will inevitably compromise the security and privacy already achieved on-chain.
For example, exchanging cryptocurrencies with tangible goods in a completely offline peer-to-peer (P2P) manner suffers from great security risks \cite{ScammersStole142021}.
%
% \junliang{For example, a peer-to-peer (P2P) trading that involves exchanging cryptocurrency for goods through direct cryptocurrency transactions between users presents security vulnerabilities}.
%
For another thing, the lack of centralized surveillance and arbitration on-chain will enhance the chances of malicious behaviors in the market and entail financial loss, which will hurt user experience.

To trade real goods with cryptocurrencies, one compromising solution is to integrate cryptocurrency payment into centralized E-commerce.
Large payment services, like Visa and PayPal, have started to allow users to exchange their cryptocurrencies with fiats based on the instant exchange rate when paying for tangible goods \cite{ExclusiveVisaMoves2021,irreraEXCLUSIVEPayPalLaunches2021}.
Although this did extend the application scope of cryptocurrencies to some extent, the trustless and anonymous properties of on-chain transactions are completely compromised.
Users need to trust and provide their private information to not only centralized E-commerce platforms as before but also these payment services.
The risk of user data leakage is still highly present, which can result in huge financial loss \cite{tripathiFinancialLossDue2020}.
%
% \junliang{Fraudulent schemes such as phishing messages are still in fact possible for causing security issues.}

%
To address the challenges mentioned above, we propose \bu{SPENDER}, not only a smart-contract-based platform that enables \bu{S}ecure and \bu{P}rivacy-Pres\bu{E}rvi\bu{N}g \bu{D}ecentralized P2P \bu{E}-comme\bu{R}ce, but also a complete paradigm for building real-world Web3 infrastructures on the blockchain.
On our platform, users
are not required to submit any personal information but their encrypted addresses, which can only be decrypted by shipping service providers.
Moreover, we introduce a bidding mechanism for shipping service providers to bid for shipping fees and delivery time, which helps build a fair, transparent, and efficient shipping market.
We build the SPENDER platform on Terra\footnote{https://www.terra.money/} which is one of the top-3 decentralized stablecoin providers.
Specifically, full functionalities of the trading system are empowered by our smart contract which we have deployed and fully tested on the Terra testnet \texttt{Bombay-12} \cite{TerraTestnets2022}.
%
% \new{Our smart contract provides full functionalities for trading system and dispute handlings.}
%
Furthermore, we develop the GUI for buyers, sellers, and shippers, as shown in Fig. \ref{fig:front_end}, and we use \texttt{Terra.js}\footnote{https://terra-money.github.io/terra.js/} to facilitate the communication between Terra wallets and the deployed smart contract.
%, as shown in Fig. \ref{fig:front_end}.

%
In summary, the contribution of this paper is two-fold.
Firstly, we build SPENDER, a fully-functional platform that supports trading real goods with cryptocurrencies in a secure, trustless, and privacy-preserving manner.
To the best of our knowledge, we are the first to develop a real decentralized E-commerce product that runs on the blockchain.
Secondly, our platform provides a complete paradigm for designing Web3 infrastructures that require both on-chain communications and off-chain interactions, where any individual that owns a digital wallet can participate.
The design concepts are fully reusable when building other DApps. 

%% file: sections/sec2-preliminary.tex
\section{Background and Related Work}
In this section, we first present some background knowledge on E-commerce, smart contract, and public-key cryptography.
Then, we analyze the pros and cons of existing centralized E-commerce platforms.
Finally, we introduce some existing Web3 platforms that support trading digital assets with cryptocurrencies and their limitations which make up part of the motivation for our work.
\subsection{E-commerce}
E-commerce, defined as any form of economic activity via an electronic network, particularly the Internet, has been considered to be essential for conducting business since the 1990s \cite{wigand1997electronic}. 
E-commerce businesses include B2C (Business-to-Consumer) such as retail sales that involve both physical goods and digital contents, B2B (Business-to-Business) such as electronic data interchange, C2C (Consumer-to-Consumer) such as auctions, and C2B (Consumer-to-Business) such as content creation.
Today, big centralized E-commerce platforms such as Amazon and Alibaba have been dominating the E-commerce market in some countries and areas\cite{winn2016secession,wei2020commerce}.
The policy and code of conduct from these platforms require the buyers and sellers to trade in a fair and lawful way.
%
% However, these platforms could be censorious and possibly exploitative. 
%

\subsection{Smart Contract}
% First para: what is smart contract
A smart contract is a transparent computer program that is automatically executed on the blockchain \cite{savelyev2017contract}.
Each smart contract is an on-chain account itself.
The execution of a smart contract is automatically triggered by the transactions sent from other accounts, and the transactions must contain certain formatted data.

%
% Second para: existing smart contracts
Ethereum \cite{buterin2014next} has become the most popular smart contract platform in terms of market cap and the available DApps and activities\cite{schar2021decentralized}.
Ethereum smart contracts are written in high-level languages such as Solidity\footnote{https://soliditylang.org/} and Vyper\footnote{https://vyper.readthedocs.io/}, and then compiled into bytecodes to be executed on the stack-based Ethereum Virtual Machine (EVM).
Other smart contract platforms include Solana\cite{yakovenko2018solana}, Polkadot\cite{wood2016polkadot}, Hyperledger Fabric\cite{androulaki2018hyperledger}, etc.

\subsection{Public-Key Cryptography}
\label{sebsec: pub-key}
As a widely used cryptographic tool for asymmetric encryption, public-key cryptography was first proposed in 1976, known as Diffie–Hellman key exchange \cite{diffieNewDirectionsCryptography1976}.
Later on, more advanced public-key schemes were invented, among which the most famous one is RSA \cite{rivest1978method}.
%
% The security of RSA algorithm largely depends on the cryptographic assumption that factoring large integers is computationally inefficient, which, however, will not be the case after quantum computers are invented \cite{shorAlgorithmsQuantumComputation1994}.
%
Other public-key encryption schemes include elliptic-curve cryptography \cite{millerUseEllipticCurves1986,koblitzEllipticCurveCryptosystems1987}, ECDSA \cite{johnsonEllipticCurveDigital2001}, and post-quantum cryptography \cite{chen2016report}.

Generally, any public-key encryption scheme can be decomposed into two processes, namely encryption $Enc(\cdot, \cdot)$ and decryption $Dec(\cdot, \cdot)$, with a pair of keys called public key $k_{pub}$ and private key $k_{pvt}$.
A secret message $m$ can be first encrypted with $k_{pub}$ as $c=Enc(m, k_{pub})$ and then decrypted with $k_{pvt}$ as $m=Dec(c, k_{pvt})$ to create an encrypted message that can only be decrypted by a specific recipient.
Alternatively, a secret message $m$ can also be first encrypted with $k_{pvt}$ as $c=Enc(m, k_{pvt})$ and then decrypted with $k_{pub}$ as $m=Dec(c, k_{pub})$ to create a non-forgeable digital signature that can be verified by the public.

%% file: figures/pipeline.tex
\begin{figure*}
    \centering
    \includegraphics[width=2\columnwidth]{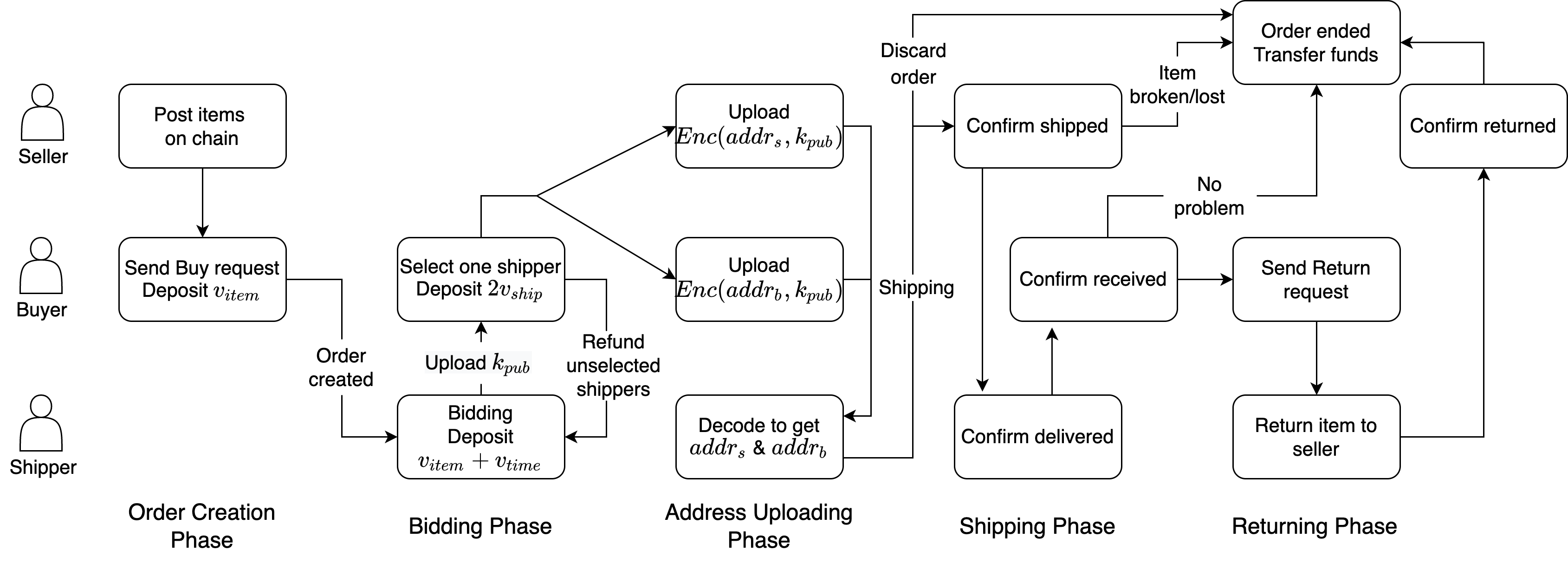}
    \caption{SPENDER System Design.}
    %\erqun{would be better if the name of the phases are all nouns, i.e. Order Creation Phase, Bidding Phase, Shipping Phase, Confirming Phase. returning phase}\erqun{It's a little difficult to tell the story when shipping is included in the third phase, which is coupled with the fourth phase. How about we put the third phase as Encrypted Address Exchaning Phase, and the fourth phase as Shipping?}\shuhao{I will update the figure as your suggestion. You can go ahead with the writing.}}
    \label{fig:pipeline}
\end{figure*}

%% file: sections/sec3-related-work.tex
% Our work is most relevant to E-commerce and Web3 applications.
% %
% In this section, we first analyze the pros and cons of existing centralized E-commerce platforms. Then we introduce some existing Web3 platforms that support trading digital assets with cryptocurrencies and their limitations which make up part of the motivation for our work.

% \begin{itemize}
%     \item Problems in E-commerce caused by center and P2P;
%     \item Blockchain as a solution for content;
%     \item Diff between our work and their work: shipper.
% \end{itemize}

%There are two popular content distribution mechanism in E-commerce.
\subsection{Centralized E-commerce Platforms}
Current E-commerce platforms such as Amazon and eBay are based on centralized Web2 techniques.
The advantage of these platforms is that they can deliver user-specific content by utilizing historical records, resolve disputes via a central control, and easily handle legitimacy issues with transparent user information.
Yet, the interest of these platforms is not always aligned with their users, which can result in privacy leakage and price discrimination.
More specifically, one company can disclose their user information to a third party for profit or present different prices to different users for the same goods by building user profiles through data.
%since the company owns all the user data.
% through big data.
%can be charged differently for different users 

%
%For example, one company may sell their user information to another company, or charge different users differently for the same goods to maximize their profit.
%
%different users may pay differently to the same good.
%
%Another is Peer-to-Peer platform such as BitTorrent~\cite{cohen2008bittorrent}, which treat every node as client and server and make them contribute to bandwidth and storage.
%
%Although this decentralized design solves the above privacy and unfair issues, it brings new problems.
%
%The users can not find the content by searching and much content infringes on copyright, which leads to legitimacy issues.

%
% The biding of shipping service providers is one of the key features we highlight in our system.
%
Moreover, the free shipping provided by Web2 platforms such as Amazon and Alibaba are established on global labor exploitation, i.e., the labor squeezing that is happening but not clearly visible to the public\cite{sathi2020transnational}. 
These platforms provide the buyers with undeniably good and secure services.
However, in order to remain competitive, businesses utilizing these platforms lead small producers and employees of shipping service providers to have few choices for their pay and working environment\cite{li2021open}.
For example, a case study for the last mile delivery work in the greater Los Angeles region shows that Amazon contributes to the deterioration of wages and working conditions\cite{10.2307/j.ctv16zjhcj.11}.
%
% Because exploitation is thought to be an effective way of lowering labor costs, and the service is targeted at reaching customers in the most speedy and least expensive way. 
%

%
Emerging Web3-based decentralized applications, as the pioneer of the future Internet, are raising attention by their potential to resolve issues, including the aforementioned ones of the Web2-based platforms.
Treiblmaier et al. \cite{treiblmaier2021impact} analyze the impacts that blockchain-based decentralized techniques can have on E-commerce.
Boersma et al. \cite{boersma2020can} discuss the potential of using blockchain-based techniques to address labor exploitation in a global context.
The decentralized E-commerce platform that we propose explores a blockchain-based system that unbinds the relationship between strong traditional E-commerce platforms and shipping services.
In our platform, shippers make bids to decide the shipping price, which helps achieve a fairer labor reward for the shippers.

\subsection{Web3 Platforms for Trading Digital Assets}
Recent work propose to leverage the blockchain mechanism~\cite{nakamoto2008bitcoin} to enable trading digital assets with cryptocurrencies in a decentralized way.
Li et al. \cite{li2020lbry} develop a blockchain-based decentralized content platform 
where users can publish, find, and pay for contents. 
This platform can also solve the discovery problem of P2P \cite{cohen2008bittorrent} by introducing a new naming system that allows users to access a global index of content metadata.
Dixit et al. \cite{dixit2021fast} propose a decentralized marketplace to enable fair IoT data trading and reliable data delivery.
Besides these digital contents, the NFTs~(Non-Fungible Token)\cite{wang2021non}, a special kind of cryptographic asset with unique identifications and public ownership traceability, have also been thriving in Web3 marketplaces.
The NFT ecosystem is also built on blockchains that serve as a public immutable transaction ledger.
%in a fault tolerant manner
%
Although these platforms can handle non-tangible goods such as e-books and music, they cannot handle tangible goods that cannot be transferred in the digital space, such as equipment, electronic devices, etc.
%Previous work~\cite{} focus on the non-tangible case  which is much more easily to handle.
%
In this paper, we introduce the shipper role into the decentralized trading framework and build a smart-contract-based platform to provide a complete system for trading tangible goods with cryptocurrencies.

%% file: sections/sec4-sys-pipe.tex
\section{System Design}
This section elaborates on the system design of SPENDER.
An illustration is shown in Fig. \ref{fig:pipeline}.
The stakeholders involved in the system are the sellers, the buyers, and the shippers.
A whole process of an order contains four major phases: the first is called the Order Creation phase, the second is the Bidding phase, the third is the Address Uploading phase, and the fourth is the Shipping phase.
An extra Returning phase will be needed if the buyer is unsatisfied with the delivered item, which will be elaborated in Sec. \ref{sec: return_phase}.
\subsection{Order Creation Phase}
The first phase, the order creation phase, starts with a seller posting an item and an obscured physical address on the smart contract.
All buyers would be able to see the item among all existing items by querying the smart contract.
An example for all visible items is shown in Fig. \ref{fig:buy_end}.
Suppose a buyer takes an interest in an item. In that case, he can order it by sending a buy request and depositing the amount of token that equals the labeled price $v_{item}$ to the smart contract, and meanwhile uploading his obscured physical address.
Then the order is created and placed on the blockchain.

\subsection{Bidding Phase}
The second phase, the bidding phase, begins when the order is placed.
A placed order is visible to all the shippers on the blockchain, who are not necessarily any postal service provider but can be any individual or organization.
From the shippers' side, the following information is available to them: (1) the item and its details posted; (2) the obscured physical addresses of both the seller and the buyer.
Then, based on the item information and the obscured addresses, multiple shippers will decide if they would like to bid for providing the shipping service.
A shipper's bid is the shipping fee $v_{ship}$ for the service of shipping the item.
%the service of shipping the item for a fee of $v_{ship}$.
%
Along with submitting the bid, the shipper also needs to deposit $v_{item}+v_{time}$ to the smart contract, which is the shipper's guarantee that he will provide delivery for the item worth $v_{item}$ within a promised delivery time.
If the delivery is not on time, the $v_{time}$ part of the deposit will not be refunded to the shipper but given to the buyer.
The buyer, upon receiving the bids, can then select a shipper to undertake the shipping task, and deposit twice the shipping fee, i.e., $2 v_{ship}$ to the smart contract.
In the meantime, the unselected shippers are refunded with their deposit $v_{item}+v_{time}$.
For the selected shipper, his deposit refund will be settled when the item is confirmed to be received in the shipping phase.
It is worth noting that with this bidding system, the price of shipping would converge to its market value.
If the shippers bid too high for the shipping fees, they would be less likely to be selected by the buyer.
If the shippers bid too low, they would not be able to profit from providing the shipping service.

\subsection{Address Uploading Phase}
The third phase, the address uploading phase, begins when a shipper is selected by the buyer.
The shipper will now request the detailed addresses of the seller and the buyer.
To achieve this, SPENDER employs the public-key cryptography schemes introduced in Sec. \ref{sebsec: pub-key}.
The shipper first posts his public key to the smart contract during the bidding phase.
The seller and the buyer then encrypt their respective detailed addresses with the public key of the shipper, and post their encrypted addresses on the smart contract.
The shipper reads the encrypted addresses from the smart contract and decrypts them with his own private key.
The specific encryption scheme can be declared by the shipper when uploading the public key, and both the encryption and decryption processes are totally off-chain.

By doing so, the detailed addresses of the seller and the buyer are not exposed to the public but only to the shipper.
This way, SPENDER preserves the address privacy of the seller and the buyer to the maximum.

\subsection{Shipping Phase}
\label{subsec: shipping_phase}
The fourth phase, the shipping phase, begins when the shipper goes to the seller's place and verifies that the item satisfies the description on the item post.
If the shipper cannot verify that the item is the same as that posted on the smart contract, he can select to discard the order.
Then the smart contract will refund the deposit $v_{item}$ to the buyer and $v_{item}+v_{time}$ to the shipper and close the order.
Otherwise, if verification goes smoothly, the shipper and the seller will both confirm that the item is shipped on the smart contract.
Next, the shipper picks up the item, goes to the buyer's address, and hands the item to the buyer for confirmation.
The shipper then confirms that the item is delivered, and the buyer, if satisfied with the item, confirms that the item is received on the smart contract.
After all the confirmations are done, the smart contract will transfer the amount of token that equals the item price, $v_{item}$ to the seller, and then transfer the shipping fee $v_{ship}$ and also the deposit $v_{item}+v_{time}$ to the shipper.
Next, the buyer can submit ratings and reviews of the order.
Finally, the order is closed automatically by the smart contract.
%
%\shuhao{Can say at last the buyer can upload ratings and reviews of the order.}

%
Besides refunding all the shipper's deposit, $v_{item}+v_{time}$, when the item is successfully delivered in the promised time, other situations can happen with the deposit.
If the item is broken or lost halfway, the shipper will take full responsibility for it, which means the deposited $v_{item}$ will not be refunded to the shipper.
In this case, the seller will be transferred $v_{item}$ from the shipper's deposit, and the buyer will also be refunded his deposit $v_{item}+2v_{ship}$.
Additionally, if the shipper delivers the item beyond the promised time, then only $v_{item}$ will be refunded to the shipper.
% \erqun{I'm confused about the $v_{item}+v_{comp}$ in the figure}\shuhao{$v_{comp}$ stands for compensation which is a fee paid to buyer if the delivery time exceeds.}\shuhao{I changed to $v_{time}$}
%

\subsection{Returning Phase}
\label{sec: return_phase}
It is not always the case that the order is carried out as described above.
Sometimes, the buyer is not satisfied with the item.
%
%\shuhao{The broken part of this paragraph is similar to the last paragraph in Sec. \ref{subsec: shipping_phase}. Also, if the item is broken, the $2v_{ship}$ will be refunded to the buyer because the shipper takes full responsibility.}
%
In this case, the buyer will need to submit a return request on the smart contract, and the shipper will return the item back to the seller. 
It is worth noting that $2 v_{ship}$ is not refunded to the buyer, which is a design to defend against malicious buyers and will be further explained in Section \ref{subsec:malicious}.
Instead, the smart contract will give two times the shipping fee, $2v_{ship}$, to the shipper.

%% file: tables/backend_message.tex
\begin{table}[htbp]
    \caption{SPENDER Back-end Messages. ``SC'' stands for Smart Contract.}
    \label{tab:back_msg}
    \centering
    \begin{tabular}{@{}llll@{}}
    \toprule
      Message Name   & Type & Permission & Funds Transfer\\
    \midrule
   Post  & Execute &  Seller & N/A \\
    %   \hline
   Buy  & Execute & Buyer & \makecell[lt]{Buyer $\xrightarrow{v_{item}}$ SC}\\
    %   \hline
   ResetPrice  & Execute & Seller& N/A \\
    %   \hline
   BidOrder  & Execute  & Shipper & \makecell[lt]{Shipper $\xrightarrow[+v_{time}]{v_{item}}$ SC}\\
    %   \hline
   ChooseBid  & Execute & Buyer & \makecell[lt]{Buyer $\xrightarrow{2v_{ship}}$ SC \\ SC $\xrightarrow[+v_{time}]{v_{item}}$ Other shippers} \\\noalign{\smallskip}
    %   \hline
   UploadAddress  & Execute & \makecell[lt]{Buyer\&\\Seller}&N/A \\
    %   \hline
   DiscardOrder & Execute & Shipper & \makecell[lt]{SC $\xrightarrow[+v_{item}]{2v_{ship}}$ Buyer \\
   SC $\xrightarrow[+v_{time}]{v_{item}}$ Shipper
   }\\\noalign{\smallskip}
   Confirm  & Execute & Everyone &\makecell[lt]{After buyer confirms \\ SC $\xrightarrow{v_{ship}}$ Buyer \\
   SC $\xrightarrow[+v_{item}]{v_{ship}}$ Shipper \\
   SC $\xrightarrow{v_{item}}$ Seller \\\noalign{\smallskip}
   If time exceeds: \\
   SC $\xrightarrow{v_{time}}$ Buyer \\\noalign{\smallskip}
   Otherwise: \\
   SC $\xrightarrow{v_{time}}$ Shipper
   }  \\\noalign{\smallskip}
    %   \hline
   ItemLossBroken  & Execute & \makecell[lt]{Buyer\&\\Shipper}
     &  \makecell[lt]{SC $\xrightarrow[+v_{item}]{2v_{ship}}$ Buyer \\
   SC $\xrightarrow{v_{time}}$ Shipper \\
   SC $\xrightarrow{v_{item}}$ Seller \\
   }  \\\noalign{\smallskip}
%   \hline
   ItemUnsatisfied  & Execute & Buyer & N/A \\
    %   \hline
   ReturnConfirm  & Execute & Seller & \makecell[lt]{SC $\xrightarrow{v_{item}}$ Buyer \\
   SC $\xrightarrow{2v_{ship}}$ Shipper
   }\\
   SubmitReview & Execute & Buyer & N/A \\
    %   \hline
   GetGoods  & Query & Everyone & N/A\\
        %   \hline
   GetOrders  & Query & Everyone & N/A\\       
%   \hline
   GetOrderDetail  & Query & Everyone & N/A\\
        %   \hline
   GetAddresses  & Query & Everyone & N/A\\
        %   \hline
   GetBalance  & Query & Everyone & N/A \\
   \bottomrule
    \end{tabular}
    %   \vspace{5pt}

\end{table}

%% file: sections/sec5-implementation.tex
\section{Implementation}
This section elaborates on the details of building the SPENDER platform, including both the back-end and the front-end parts.
\subsection{Back-end}
Back-end design consists of all the logistics in smart contract interactions. 
As a smart contract is generally a message-passing system, there are mainly two types of messages to be implemented, namely execution messages and query messages.
Execution messages are messages that will change the internal states of a smart contract and will be charged for transaction fees as rewards paid to blockchain miners.
Query messages, on the other hand, are messages to get the information about the internal states of a smart contract and are free of charge.
TABLE \ref{tab:back_msg} lists all the messages implemented in our smart contract, accompanied by the permissions and funds transferred during execution.

Our smart contract is first written in Rust \cite{matsakis2014rust} with CosmWasm Software Development Kit \cite{CosmWasm2022}, and then compiled into WebAssembly \cite{RossbergWebAssemblyCoreSpecification} code using \texttt{rust-optimizer} \cite{CosmWasmOptimizingCompiler2022}.
Afterwards, the WebAssembly code is deployed directly onto Terra testnet \texttt{Bombay-12} \cite{TerraTestnets2022} for further testing.
\subsection{Front-end}
As shown in Fig. \ref{fig:front_end}, the front-end of SPENDER consists of three specific Graphical User Interfaces (GUI) for sellers, buyers, and shippers, respectively.
The GUI is written in JavaScript.
To connect the GUI to Terra's wallet and smart contract on the testnet, we adopt the official toolkit \texttt{Terra.js} \cite{TerramoneyTerraJs2022} which reads and converts messages to \texttt{JSON} format.
To avoid copyright issues, we use the stock images collected from Shutterstock\footnote{https://www.shutterstock.com} with some dummy texts.

%% file: sections/sec6-adv-features.tex
\section{Advantageous Features}
In this section, we discuss seven advantageous features of SPENDER, the first decentralized P2P E-commerce platform in the world.
\subsection{Malicious Behaviors Resistance}
\label{subsec:malicious}
SPENDER has inborn robustness towards some kinds of malicious behaviors.
Besides, SPENDER guarantees that the other malicious behaviors which SPENDER cannot perfectly defend against have little possibility of success.
% \shuhao{may need revise.}
%
% \old{The argument is that these cases are also difficult to deal with in traditional E-commerce, and therefore they should not be viewed as disadvantages of SPENDER.}
% \noindent\textbf{Malicious seller alone}\quad 
\subsubsection{Malicious seller alone}
If a seller sells fake or broken items deliberately to trick the buyer's money, the shipper can simply discard the order upon checking the item at the seller's.
This way, our system can defend against the malicious behavior of sellers towards buyers.
Plus, our system can also defend against malicious behavior of sellers toward shippers, i.e., posting inaccurate item information and making a shipper travel in vain.
For one thing, a seller cannot target a particular shipper because the shippers are chosen by the buyers.
For another, the seller will have to pay the gas fee to post the item, which can be prohibitive for such a prank without any profits. 
%

%
% \noindent\textbf{Malicious buyer alone}\quad
\subsubsection{Malicious buyer alone}
A buyer can attack a shipper by deliberately returning the item.
However, by our design, the deposited $2 v_{ship}$ will not be refunded to the buyer if the item is returned.
Therefore, buyers have little motivation to conduct such behavior.
%

%
% \noindent\textbf{Malicious shipper alone}\quad
\subsubsection{Malicious shipper alone}

% \shuhao{This paragraph should be organized as 2-1-4-3. Our system can... $\rightarrow$ shipper's duty, they can only earn by perfect delivery $\rightarrow$ We also have rating.}
%
Our system can also discourage malicious shippers.
For SPENDER to function well, it is desired that the shippers carry out their duty, i.e., they need to verify the item at the sellers and deliver the item without breaking or losing it.
Therefore, in SPENDER, the shippers can only earn profits by perfect delivery, which will not be achieved if an order fails.
Plus, the rating system is also important for encouraging the shippers to carry out their duty.
%

% \noindent\textbf{Malicious buyer and shipper}\quad 
\subsubsection{Malicious buyer and shipper}

Our system can prevent a buyer and a shipper from being complicit in attacking a seller because they lack incentives.
Suppose the buyer selects the shipper who is complicit. 
Even if the buyer starts a malicious return, or the shipper makes damage to the item intentionally, the shipper and the buyer will have no benefits.
This is because the shipping fee and return fee $2v_{ship}$ are from the buyer, as shown in Fig. \ref{fig:pipeline}, and the shipper has to pay for the compensation.
%

%
% \noindent\textbf{Malicious seller and shipper}\quad 
\subsubsection{Malicious seller and shipper}

This situation would have the potential to cause a shipping fee scam targeting a buyer if a seller and a shipper are complicit.
The seller and shipper have no motivation to damage the item as the compensation is paid by the shipper.
A possible fraud is the malicious behavior that forces the buyer to return the item, considering that the buyer shall provide the return fee.
The seller could intentionally put a defective or wrong item into the package, and the shipper skips the verification when picking up the item.
This malicious behavior will make the buyer have to make a return so that the shipper will take the double shipping fee from the buyer.
However, this can happen only if the buyer picks up the malicious shipper from a list of shippers. 
Though it is not possible to completely avoid the situation if the shipper makes a regular bid, still an abnormally low bid should definitely alarm the buyer.
%

%
% \noindent\textbf{Malicious buyer and seller}\quad 
\subsubsection{Malicious buyer and seller}
With this being the case, a malicious buyer and seller could make some phishing orders to obtain the compensation by fraud, considering that the compensation is paid by the shipper.
However, there is a chance that the shippers can protect themselves from phishing orders under the bidding system.
Suppose a seller posts a random item selling for a ridiculous price on purpose, and a buyer who is complicit buys the item to make an order.
Theoretically, if the item arrives damaged, the shipping service provider has to pay the compensation.
However, shippers can choose not to make a bid and thus refuse to provide the service if they see an abnormal price.
Moreover, the review system also helps shippers to detect phishing orders.
%

%
% \noindent\textbf{Malicious buyer, seller, and shipper}\quad 
\subsubsection{Malicious buyer, seller, and shipper}
Under the Web2 environment, a group of malicious buyers, sellers, and shippers may engage in arbitrage to obtain money from a Web2 platform. 
For example, a fraud group can do a fake business to make a profit by taking advantage of the seller support promotion from a platform\cite{xu2015commerce}.
This problem has not been evident for Web3 yet but merits further consideration.
Another challenge is E-commerce brushing\cite{jin2019brush}, also known as click farm, which targets to cheat the review system.
A seller may create fake orders to be completed with the help of some fake buyers and shippers to boost the ratings.
Nevertheless, E-commerce brushing on SPENDER comes at a cost. 
A gas fee is charged for any transaction, and all the details of the fake orders are publicly visible, which creates the possibility of distinguishing the fake orders for the users.
As the review system remains to be an essential part of Web3, it is worth proposing new methods for solving the problem in the future.

\subsection{Permissionlessness and Compatibility}
SPENDER is designed to allow any individual to participate and resist censorship, but this doesn't mean SPENDER will invade and threaten the centralized E-commerce market.
Instead, SPENDER provides perfect compatibility to traditional E-commerce platforms and shipping service providers, allowing them to participate as single sellers and single shippers.
In this way, SPENDER becomes an additional choice for the shareholders in the traditional E-commerce market.
Meanwhile, SPENDER allows individual sellers and shipping service providers to enter the market and contribute to market efficiency.
\subsection{Minimum Privacy Leakage}
One of the minimum requirements for a complete delivery is that the shipper should know the addresses of both the buyer and the seller.
In SPENDER, the public-key cryptography guarantees that only the shipper knows the exact addresses of the buyer and the seller.
Except for the addresses, no private information is required.
Moreover, the randomness of shipper bidding ensures the sparsity of private information distribution among shippers.
Therefore, we can claim that SPENDER achieves the minimum privacy leakage in a decentralized E-commerce system.
\subsection{Financial Security}
Centralized E-commerce platforms usually function as cash-keepers before orders are finished.
Money paid by buyers is locked in these platforms for financial security.
Even if these cash-keepers are assumed to be one-hundred percent trustful and safe, they can be completely replaced by a smart contract on the blockchain.
A smart contract is itself an on-chain account that can keep cryptocurrencies.
The money kept in the smart contract cannot be moved or used by any individual except the built-in automatic program, which is totally transparent.
As long as the smart contract is well-written and the blockchain itself is not attacked, the locked cryptocurrencies are safe.
Terra adopts the robust programming language Rust as the backbone to resolve the vulnerabilities in smart contracts written in Solidity \cite{pariziEmpiricalVulnerabilityAnalysis2018}.
At the same time, Terra runs a Proof-of-Stake (PoS) consensus protocol to enhance the security of its blockchain \cite{kereiakes2019terra}.

\subsection{Price Stability}
One outstanding problem in cryptocurrency trading is the volatility of the price.
No seller is willing to see the price of their goods varies with the cryptocurrency market, which is out of their control.
This is why we choose to build SPENDER on the Terra ecosystem.

Terra is the world's top-3 provider of stablecoins, i.e., cryptocurrencies whose price is always bound with fiats, unlike BTC and ETH whose price fluctuates a lot.
For example, one TerraUSD (a.k.a. UST) always has the same value as one USD.
Moreover, Terra also provides many other stablecoins that are bound with other fiats in the world, such as TerraKRW (a.k.a. KRT), which has a value equal to KRW.
This stablecoin ecosystem guarantees the price stability of SPENDER and will enable international trading if SPENDER is integrated with Decentralized Exchanges (DEXs) for currency conversion in the future.
\subsection{Transparent Review System}
Centralized E-commerce platforms are commonly equipped with review systems, in which buyers can write reviews and rate their shopping experience for their orders.
A good review system can help buyers discriminate good sellers from bad sellers and improve buyers' shopping experience.

However, there is no guarantee that the statistics of the review systems in centralized E-commerce are real because the systems are not transparent.
By contrast, in SPENDER, since all the order details are recorded on the blockchain, it is very easy to compute the ratings and other statistics for each participant in a transparent manner.
Two examples are the ratio of perfect delivery for shippers and the ratio of satisfied goods for sellers.
In this way, the review system in SPENDER also encourages sellers and shippers to stay at the same wallet address and accumulate good reviews, which achieves the same effect as flagship stores and shipping services in centralized E-commerce.
\subsection{Illegal Goods Prevention}
Legitimacy has always been a controversial topic in decentralized systems.
One specific concern is that illegal transactions may go wild without centralized regulations.
However, SPENDER is naturally designed to avoid selling illegal goods.
Specifically, if sellers who are selling illegal goods are spotted, law enforcement agencies can easily conduct entrapment to find and arrest the illegal sellers by acting as the buyer and the shipper simultaneously.
This is made possible by the bidding mechanism in SPENDER, in which buyers can select specific shippers, and shippers can know the addresses of sellers.

%% file: sections/sec7-future-potentials.tex
\section{Future Potential}
In this section, we list three promising future directions in improving the SPENDER platform.

\subsection{Decentralized Arbitration via DAO}

Disputes can happen wherever the physical world is involved.
Same as in traditional E-commerce, real-world consensus problems also exist in SPENDER.
For example, shippers may deny they have broken the items and refuse to compensate.
Sellers may sell fake goods and insist that they are genuine.
In such scenarios, arbitration is needed to handle the disputes.
However, without centralized authorities, arbitration can only be executed in a decentralized way.

Decentralized Autonomous Organizations (DAO) are organizations ruled by automatic and transparent smart contracts without any central authorities \cite{beckGovernanceBlockchainEconomy2018}.
Therefore, DAOs are perfect for decentralized governance, and arbitration \cite{dialloEGovDAOBetterGovernment2018}.
When a dispute happens, each involved party can upload related proofs onto the blockchain to ask for votes via a DAO.
The final decision will be achieved democratically and executed automatically by the smart contract. 
\subsection{Access Control}

In the current design of SPENDER, every detail of the orders is transparent.
However, as also mentioned in \cite{treiblmaier2021impact}, legislation in most countries may regulate the accessibility and usage of users' sensitive data.
Therefore, it is necessary to have an access control system in SPENDER to manage the access permissions.
Several pioneering work have proposed solutions to on-chain access control using smart contracts \cite{steichenBlockchainBasedDecentralizedAccess2018,guoMultiAuthorityAttributeBasedAccess2019}.
We plan to design an access control system based on these existing solutions in the future.

\subsection{On-chain Recommendation Services}

Personalized recommendation services have become an essential part of traditional E-commerce platforms, aiming to alleviate information overload and cater to users' different demands.
While there are numerous studies and applications on recommender systems in Web2 E-commerce platforms~\cite{Wide_Deep, youtube_rec, Guo2017Apriori, HWANGBO201894, Wu2021MultiFR, Ma2021knowledge}, there are insufficient studies on deploying recommendation services on Web3 decentralized platforms.
Therefore, we also plan to establish a recommender system for SPENDER in the future.
%

% \old{In comparison to traditional recommender systems for Web2 E-commerce, one of the challenges in deploying on-chain recommendation services in SPENDER is the lack of a stable identifier for each consumer.
% %
% For preserving users' privacy, the addresses used by the consumers on-chain may vary, and some addresses may even become inactive following a single business transaction.
% %
% Thus, we intend to develop our recommender systems based on the similarity of transaction behavior, i.e., the community-based recommendation.
% %
% When the keywords queried by a user are detected on-chain, the recommendation service can first group the wallets with similar transaction behavior into communities based on their historical purchase records and then offer a personalized ranking list for the user.
% %
% By doing this, we can enrich the data for those wallets with fewer transaction histories during training the recommendation models, thus alleviating the ``cold-start'' problem~\cite{/BobadillaOHB12cold_start} to some extent.}
% \erqun{We are not here to explain the challenge. I think we should use one sentence to summarize all, and focus on potential.}
We intend to develop our recommender system based on either the purchase records of each wallet or the purchase records of the wallet groups that share similar transaction behavior.
Additionally, we would also like to allow third-party recommendation services on-chain.
As a result, consumers will have the option of selecting the recommendation service provider they prefer.
%and make their own decisions in purchasing.
%
Moreover, these third-party recommendation service providers can charge fees for better recommendation results, forming a competitive yet efficient recommendation market.

%% file: sections/sec8-conclusion.tex
\section{Conclusion}
In this paper, we present SPENDER, a decentralized P2P E-commerce platform 
%that provides a series of design concepts for designing real-world decentralized applications.
%
%The objective of SPENDER is 
%which propels the forefront in designing a Web3 E-commerce system 
that simultaneously preserves privacy, transparency, and information security.
Specially, we adopt public-key cryptography to encrypt users' physical addresses to prevent the exposure of the shared addresses to unrelated users, and we utilize the smart-contract-based deposit mechanism to ensure financial security.
Moreover, we introduce a shipping fee bidding mechanism to attain a fair shipping market and encourage shippers to provide better services.
We have built SPENDER using Terra CosmWasm architecture and have fully tested its functionalities on the Terra testnet.
SPENDER shows great advantage in solving the existing privacy-related issues in traditional E-commerce platforms and provides a practical paradigm for designing Web3 applications that require both on-chain communications and off-chain interactions.
%